\documentclass[proceedings]{JHEP3}

\PrHEP{PrHEP hep2001}                   
\conference{International Europhysics Conference on HEP}                

\title{The cosmological information on neutrino mixing}

\author{\speaker{Pasquale Di Bari} \\                     
   Deutsches Elektronen-Synchrotron DESY, 22603 Hamburg, Germany \\  
   E-mail: \email{dibari@mail.desy.de}}                       

\abstract{Cosmology provides interesting
information on neutrino mixing models with sterile neutrinos.
In this case non standard BBN effects can be relevant. We show how 
the recent measurement of the baryon content from the observations of CMB 
anisotropies together with the primordial nuclear abundances 
measurements can be used to constrain them. In particular four
neutrino mixing models are potentially at variance with the cosmological observations. 
We also discuss the possible scenarios from future experiments.
}

\begin{document}
 \section{Constraints on non standard BBN models}

     Evidences for neutrino mixing have been found both in atmospheric neutrino and
     in solar neutrino data and together they provide the first solid indication
     of new physics in Earth experiments. Hints of new physics were already present also  
     in cosmology. For example the measurement of a total matter content 
     larger than the baryonic one pointing to the existence of a new type of
      matter. The baryonic content was inferred from the model of standard 
      Big Bang Nucleosynthesis (SBBN) using the primordial nuclear abundances 
      measurements. The recent high resolution observations of CMB anisotropies 
      have confirmed the existence of acoustic peaks in the power spectrum as 
      predicted by earlier theoretical studies that also relate their features 
     to the values of the cosmological parameters. In this way it has already been possible
     to get a measurement of the baryon content independently on SBBN. This measurement
     relies on some minimal assumptions, but it is quite stable respect to
     the relaxation of many of them. The agreement, within the errors, with the value
     inferred from SBBN, strongly supports the existence of non baryonic dark
     matter and can be used to constrain a large variety of non standard BBN 
     effects. We are interested in those non standard BBN effects that arise 
     in neutrino mixing models and we want to see which kind of information 
     can be extracted from current cosmological observations on them
     \footnote{For details see \cite{DiBari} and references therein.}. 
	From the observation of CMB anisotropies, 
      the BOOMERanG and DASI experiments find for the baryon contribution
     to the energy density $(\Omega_b\,h^2)^{CMB}=0.022^{+0.004}_{-0.003}$, 
     that implies a value of baryon to photon ratio at the time of recombination
     (in units of $10^{-10}$) $\eta^{CMB}=6.0^{+1.1}_{-0.8}$.  
      This value can be used in the calculation
     of primordial nuclear abundances within BBN models. In the case of SBBN 
     the knowledge of this value makes possible to predict all the primordial nuclear 
     abundances and in particular the helium-4 ($Y_p$) and the deuterium abundances 
     that are measured in the astrophysical observations with reasonable reliability 
     and precision. The comparison between the SBBN predictions and the measured values 
     gives a very good agreement for `high values' of $Y_p$,
     while there is a $4\,\sigma$ discrepancy in the case of `low values' (SBBN `crisis').  
    Thus at the moment the simplest interpretation of the data is a confirmation of SBBN. 
    However one can study how large deviations from SBBN  can be if non standard
    BBN effects are considered. We are interested in {\em two different
    classes of non standard BBN effects}. The {\em first} one is the presence of 
    extra energy density degrees of freedom that can be expressed through the {\em extra 
    number of neutrino species} $\Delta N_{\nu}^{\rho}$. The {\em second} one
    is a deviation of electron neutrino and anti-neutrino distributions from the
    SBBN assumption of a Fermi-Dirac distribution with zero chemical potential.
     These two classes account for a large variety of non standard effects 
     from theories beyond  the Standard Model of fundamental interactions
\footnote{Another class of possible non standard effects, not considered here,
is for example that one accounting for clumps or holes in baryon number density.}.
        The presence of a non zero $\Delta N_{\nu}^{\rho}$ affects the standard BBN
      prediction on $Y_p$ by a quantity $\Delta Y_p\simeq 0.0137\,\Delta N_{\nu}^{\rho}$.
     The effect on the Deuterium abundance is double. There is an indirect effect due to the           variation of $Y_p$ itself, but there is also a direct effect due to change of the rate expansion induced by $\Delta N_{\nu}^{\rho}$. 
	The distortions of electron neutrino distributions can influence only the
      frozen value of the neutron to proton ratio at    
     a temperature around $0.75\,{\rm MeV}$. When the nuclear abundances start to get
     synthesized at much lower temperatures around $\sim 0.065\,{\rm MeV}$, 
the electron neutrinos and anti-neutrinos are fully decoupled and they do not play any role. 
      Thus the effect of distortions of electron neutrino distributions 
      have a direct effect only on $Y_p$. One can define a total effective number
      of extra neutrino species $\Delta N_{\nu}^{\rm tot}\equiv
        [Y_p(\eta^{CMB},\Delta N_{\nu}^{\rho},\delta f_{\nu_e,\bar{\nu}_e})-
        Y_p^{SBBN}(\eta^{CMB})]/0.0137 $. Note that in general the distortions of 
       the electron neutrino distribution are time dependent during the phase of neutron to                                                           proton ratio freeze-out. Note also that in general
       they can also give a contribution to $\Delta N_{\nu}^{\rho}$. Thus in general
       $\Delta N_{\nu}^{\rho}$ could change with time during BBN period. However in the
       models we will consider this effect is negligible and the value of 
       $\Delta N_{\nu}^{\rho}$ can be considered constant during the BBN epoch.
      A comparison of the measured value $Y^{\rm exp}_p$ with $Y_p^{SBBN}$ can thus be
       translated in terms of $\Delta N_{\nu}^{\rm tot}$ and we find, conservatively
       at $3\,\sigma$, $-1.8<\Delta N_{\nu}^{\rm tot}<0.3$.  The possibility to 
       distinguish separately the value of $\Delta N_{\nu}^{\rho}$ is given by a 
       comparison between the predicted and the measured value of deuterium abundance.
        In this way one has a correspondence $(\Delta Y_p,\Delta (D/H))\leftrightarrow 
       (\Delta N_{\nu}^{\rm tot},\Delta N_{\nu}^{\rho})$, where $\Delta$ 
       indicates a variation compared to the SBBN case. This change of variables
       is not necessary in principle but makes possible to quantify in a similar
       way two different effects.  Moreover the bi-parametrical nature of the
       non standard BBN effects that we are considering (we are dealing with 
       $\eta$ as an independently measured quantity, like the neutron life time)
       is emphasized and this would hold considering also more than two
       primordial nuclear abundances (for example if one could rely primordial
       lithium-7 estimations). 

    \section{One or more sterile neutrino flavors ?}

         It is known since long time that the early Universe is 
       sensitive to a mixing of the ordinary neutrinos with new possible sterile
       neutrinos. The simplest case is to consider a mixing 
       between one ordinary and one sterile neutrino in isolation. 
       In such a situation the mixing is described by just two parameters, 
        the difference of squared masses $\delta m^2$ and $\sin^2 2\theta_0$ 
        ($\theta_0$ is the vacuum mixing angle). The task is then to calculate
        $\Delta N_{\nu}^{\rho}$ and $\Delta N_{\nu}^{\rm tot}$ for any given 
        $(\delta m^2,\sin^2\,2\theta_0)$.
         For values $\sin^2\,2\theta_0\gtrsim 10^{-4}$ it has been found 
         that one always gets a positive $\Delta N_{\nu}^{\rm tot}$. 
         For values $\sin^2 2\theta_0\lesssim 10^{-4}$ and negative values of $\delta m^2$
	  the situation is quite different because a large neutrino asymmetry 
        is generated \cite{ftv}. In the case that the ordinary neutrino is an 
         electron neutrino and if the asymmetry is positive, then the effect on BBN is 
        that $\Delta N_{\nu}^{\rm tot}$ can be negative and this could be a way 
        to reconcile standard BBN with low values of $Y_p$.
        The neutrino asymmetry generation 
        appears as a special possibility within the simple case of two neutrino mixing.
        However when multiflavor mixing is considered, the much larger choice  	   
        of mixing parameters makes more natural to have the right conditions   
	   of neutrino asymmetry generation being satisfied for some of the many
        possible sub mixings. In this case what is important is that the generated
        neutrino asymmetry suppresses the neutrino oscillations and thus also a
        sterile neutrino production \cite{fv0}. Therefore the constraints
 	  found in the two neutrino mixing can be evaded when this is embedded in a 
        multi-flavor scenario. The neutrino asymmetry generation occurs for mixing angles
        much lower than those tested in the Earth experiments and so cosmology 
        can provide a complementary information. The lower limit is 
        given by the condition of adiabaticity. This limit has been studied in detail in 
        \cite{ropa} where it has been found that the MSW effect, 
        responsible for the generation of the asymmetry, is adiabatic
        for $\sin^2 2\theta_0 \gtrsim 10^{-9}\,({\rm eV}^2/|\delta m^2|)^{1/4}$. 
     The simplest realistic case of multiflavor mixing involving sterile neutrinos 
    is a four neutrino mixing between one sterile neutrino and the three ordinary
    neutrinos. This is the minimal way to explain all existing experimental neutrino
    mixing evidences, the solar neutrino data, the atmospheric neutrino data and the 
    short baseline LSND experiment results. This last experiment requires the existence      
    of a $1\,{\rm eV}^2$ scale for $\delta m^2$, much different from those that
    one infers for the solar and atmospheric neutrino data and thus a fourth neutrino mass                                 and flavor eigenstate  are necessary to built a neutrino mixing model consistent with all experimental data. There are {\em two different ways} to achieve this result. 
The {\em first} one is just by adding the sterile neutrino flavor as a sort of 
    `perturbation' to the three neutrino mixing matrix. In this way it almost coincides 
    with the fourth mass eigenstate separated by the LSND gap from the
    other three (`3+1' models) and their mixing is such that the sterile neutrino flavor 
    is brought to thermal equilibrium or very close to it in a way that 
     $\Delta N_{\nu}^{\rho}\simeq 1$.
    The condition for a significant electron neutrino asymmetry generation able to 
    modify the BBN predictions are not realized and therefore the final result is 
    $\Delta N_{\nu}^{\rm tot}\simeq 1$, incompatible with the cosmological bound.
	The {\em second} way represents a drastic change
    compared to three neutrino mixing models. In this case the four mass eigenstates are 
    split in two couples separated by the LSND gap. The simplest realization of this class
    of models (`2+2' models) is to have a full correspondence between each pair of mass 
    eigenstates with a pair of neutrino flavors 
    \footnote{It corresponds to have a 2x2 block factorization of the mixing matrix.}. 
     In this simplified picture,
    the solar neutrino data are described by oscillations of $\nu_e$ into $\nu_{\mu}$ or
    $\nu_s$ while the atmospheric neutrino data by oscillations of $\nu_{\mu}$ 
     into $\nu_{\tau}$ or $\nu_s$. A small mixing between the two pairs has to be
    included to explain the LSND results but in first approximation one has four limit                                                                                  cases in which solar or atmospheric neutrino data (not both) are explained by oscillations 
    of active neutrinos into sterile neutrinos. From a cosmological point of view
    one has to worry for those cases that include active to sterile neutrino oscillations 
     with large mixing angles: $\nu_{\mu}$ to $\nu_s$ to explain atmospheric neutrino data
     or the LMA solution $\nu_e$ to $\nu_s$ to explain solar neutrino data.  
     One can again slightly perturb these 4 limit cases in order to get a small mixing
     between the two couples such to generate a neutrino asymmetry that suppresses
     the sterile neutrino production \cite{bell}. However the most recent analysis for
     the atmospheric and solar neutrino data  taken separately,  
     disfavor approximately at $3\,\sigma$ the possibility of pure active to 
     sterile neutrino oscillations. In a joint four
    neutrino mixing analysis of solar and atmospheric neutrino data, the possibility 
     to have pure active to sterile neutrino oscillations in one of the two cases                                                    is only slightly disfavored but the best fit
     is obtained for those models in which the muon neutrinos are converted into a mixture
     of sterile and tau neutrinos in the atmospheric case and the electron neutrinos 
     into a mixture of ordinary (muon and tauon) and sterile neutrinos in the solar case
      \cite{gonzalez}. 
        In this situation the sterile neutrino flavor is brought to thermal equilibrium
       and moreover a neutrino asymmetry generation able to suppress the sterile                   neutrino production or to yield a negative contribution to $\Delta N_{\nu}^{\rm tot}$
     is not possible. Therefore, like for 3+1 models, one gets again 
    $\Delta N_{\nu}^{\rm tot}\simeq 1$, at variance with the cosmological bound. 
    Future data should be able to constraint even more the possibility of pure active to sterile neutrino oscillations and thus the possibility to have a neutrino asymmetry generation in four              neutrino mixing models. If this will be the case then the possibility to include the 
     LSND results in a consistent four neutrino mixing picture not conflicting with the    cosmological observations relies on the assumption that large neutrino asymmetries 
were generated at temperatures much above $10 {\rm MeV}$ (`pre-existing' 
neutrino asymmetries). This would be an important 
piece of information for models of baryo-leptogenesis.
  Another possibility is to assume the existence of a second sterile neutrino    
   flavor mixed with the ordinary ones with the appropriate mixing 
   parameters  such that a positive electron neutrino asymmetry is generated. This would 
   both result in a negative contribution to $\Delta N_{\nu}^{\rm tot}$  and at the same time                  would halve the sterile neutrino  production in a way to have 
$\Delta N_{\nu}^{\rho}\simeq 0.5$ and $\Delta N_{\nu}^{\rm tot}<0.3$ . 
In this way it would be possible to reconcile the LSND result with cosmology. 
For this to happen the electron neutrino flavor should be mainly `contained' 
in the $1\,{\rm eV}$ mass eigenstates. If this is true the next generation of 
$\beta\beta_{0\nu}$ experiments could be able, under some 
conditions, to detect a neutrino mass. 
Neutrino mixing models with more than one sterile neutrino flavors arise, for example, 
in theories with large extra dimensions. In this case the role of
sterile neutrinos is played by the Kaluza Klein states of one fermion
living on the bulk and mixed with the ordinary neutrino on the brane. 
Unfortunately all models proposed up to now do not seem to solve the problems
of four neutrino mixing models.

    \section{Possible future scenarios}

     In future years both cosmological observations and neutrino mixing experiments
     will provide many new data. From the CMB power spectrum the Planck satellite should be 
     able to measure the baryon content with a $1\%$ error
     and at the same to detect $\Delta N_{\nu}^{\rho}$ with a realistic error of 0.1. 
All these new data will be  able to constraint very strongly the possibility of 
non standard BBN effects from active-sterile  neutrino mixing. We can list
the possible scenarios, assuming that pure active to sterile neutrino oscillations 
will be definitively ruled out both in solar and atmospheric neutrino data, that means
to exclude the case that four neutrino mixing models are compatible with cosmology 
without assuming pre-existing large neutrino asymmetries. 
1) The LSND result is confirmed
but {\em non} SBBN effects are not found, even though expected. 
A likely explanation would be the assumption of pre-existing
large neutrino asymmetries. 
2) LSND is confirmed but {\em non} SBBN effects are discovered and
this means ($\Delta N_{\nu}^{\rho},\Delta N_{\nu}^{\rm tot})\neq (0,0)$. 
A result $\Delta N_{\nu}^{\rm tot}<\Delta N_{\nu}^{\rho}$ would be a specific 
signature of electron neutrino asymmetry generation and thus of active-sterile neutrino mixing.  
 In this case one can conclude that there are additional sterile neutrino flavors 
other than that one suggested by the LSND experiment. If one can attribute 
the effects to neutrino mixing, then one can also remarkably conclude that 
pre-existing  neutrino asymmetries were small.
3) {\em Non} SBBN effects are found but LSND is disproved. 
      In this case the problem to attribute the effects to active-sterile
      neutrino mixing would be crucial. As we said a result
       $\Delta N_{\nu}^{\rm tot}<\Delta N_{\nu}^{\rho}$ would represent a good signature.  
       The possible formation of {\em neutrino domains} 
       associated to the neutrino asymmetry generation \cite{domains} 
       would have effects \cite{dario} that could provide other kind of signatures.
4) LSND is disproved and cosmological observations 
do not find any {\em non} SBBN effect. In this case a three ordinary
neutrino mixing will be enough to explain all the data and it will be challenging to 
search for cosmological effects of it. 

\acknowledgments
P. Di Bari is an Alexander von Humboldt Foundation fellow. He thanks 
M. Lusignoli and L. Mersini for valuable discussions.



\end{document}